\documentclass[aps,prl,superscriptaddress,twocolumn]{revtex4}
\usepackage{bm}
\usepackage{graphicx,amsmath,latexsym,amssymb}
\usepackage{color}
\usepackage{dcolumn}

\begin{document}

\title{Casimir Momentum of a Chiral Molecule in a Magnetic Field }

\author{M. Donaire }
%\email{manuel.donaire@grenoble.cnrs.fr}

\affiliation{Universit\'{e} Grenoble 1/CNRS, LPMMC UMR 5493, B.P.
166, 38042 Grenoble, France}

\affiliation{LNCMI, UPR 3228 CNRS/INSA/UJF Grenoble 1/UPS, Toulouse
\& Grenoble, France }

\author{B.A. van Tiggelen}

\affiliation{Universit\'{e} Grenoble 1/CNRS, LPMMC UMR 5493, B.P.
166, 38042 Grenoble, France}

\author{G.L.J.A. Rikken}

\affiliation{LNCMI, UPR 3228 CNRS/INSA/UJF Grenoble 1/UPS, Toulouse
\& Grenoble, France }

\date{\today}

\begin{abstract}
In a classical description, a neutral, polarizable  object acquires a kinetic momentum when exposed to crossed electric and magnetic fields. 
In the presence of only a magnetic field no such momentum exists classically, although it is symmetry-allowed for an object lacking mirror symmetry. We perform a full QED 
calculation to show that the quantum vacuum coupled to a chiral molecule provides it with a kinetic "Casimir" momentum directed along the magnetic field, and proportional to 
its molecular rotatory power and to the fine structure constant.
\end{abstract}


\maketitle
The notion that electromagnetic quantum vacuum fluctuations contain energy is widely accepted. When coupled to matter they give rise to Van-der-Waals forces between
neutral objects and Lamb-shifts of energy levels in atoms \cite{Milonni}. The attraction between metallic plates is undoubtedly the most convincing proof 
for the existence
of Casimir energy \cite{Casimir}. The empty quantum vacuum is Lorentz-invariant with a diverging energy density with frequency spectrum $\omega^3$ and zero momentum. 
The regularization of this divergence is still an issue in the context of the cosmological constant 
problem \cite{Jaffe}. In all other cases the UV catastrophe is not essential when binding energies and forces are calculated in a dielectric medium \cite{Manuel}. 

The possibility of the quantum vacuum to contribute to the momentum of matter was first raised in Ref.~\cite{Feigel}. For a macroscopic object exposed to 
crossed electric and magnetic fields, $\mathbf{E}_{0}$ and $\mathbf{B}_{0}$ respectively, a QED correction was obtained in addition to the classical
Abraham momentum density, $(\epsilon_{r}-1) (\mathbf{E}_0\wedge \mathbf{B}_0)$, which emerges from Maxwell's equations and is already controversial by itself as concerning the prefactor.
Being an observable quantity,
the UV divergence of the momentum obtained in Ref.~\cite{Feigel} poses a real problem if only because the end result depends crucially on an heuristic cut-off. 
In Refs.~\cite{Kawka,vT3} 
two of us have shown that the divergence can be handled in a full quantum theory that goes beyond the dipole approximation. Divergencies were 
seen to disappear by mass regularization and a finite but small Casimir momentum remained that is a factor $\alpha^2$ smaller than the classical effect. In the 
quantum theory it is also clear that two contributions of the vacuum field exist to the total momentum. One is the familiar sum 
$\sum_{\mathbf{k},\mathbf{\epsilon}} \hbar \mathbf{k} (a^{\dagger}_{\mathbf{k}\mathbf{\epsilon}}a_{\mathbf{k}\mathbf{\epsilon}}+1/2)$
associated with photons with momentum $\hbar \mathbf{k}$, and often referred to as the "transverse" component \cite{Cohen}. The second more subtle 
contribution, $\sum_{i} q_{i} \mathbf{A}_{\perp}(\mathbf{r}_{i})$, stems from the transverse electromagnetic gauge field coupled to electric charges $\{q_{i}\}$ and 
is manifestly gauge invariant. This is referred to as "longitudinal" component since it can be written as the integral over the vector product of the longitudinal electric field and 
the magnetic field, $\epsilon_{0}\int\textrm{d}^{3}r\mathbf{E}_{\parallel}\wedge\mathbf{B}$ \cite{Cohen}. The longitudinal component is found to dominate the Casimir momentum.

In this work we apply the same formalism as in Refs.~\cite{Kawka,vT3} to address the Casimir momentum of a quantum object subject to only a magnetic field. 
The latter being a pseudo-vector, symmetry imposes the object to be chiral, i.e. we anticipate the relation $\mathbf{P}^{\textrm{Cas}}= e g\mathbf{B}_0$ with $g$ some 
pseudo-scalar quantifying the broken mirror symmetry of the quantum object. Important is that no classical contribution has been reported, 
so that the proposed contribution from the quantum vacuum would
constitute 100$\%$ of the effect. Our theory is nonrelativistic, and gives a finite magneto-chiral Casimir momentum. We find $g$ to be 
proportional to the product of the static rotatory power of the molecule and the fine structure constant.

We propose the simplest molecular model that exhibits all necessary features to leading order in perturbation theory: 
broken mirror symmetry, Zeeman splitting of energy levels and coupling to the
quantum vacuum, but we neglect relativistic effects. In our model the optical activity of the molecule is determined by a single chromophoric electron within a 
 chiral object which is further simplified to be a two-particle system in which the chromophoric electron of 
charge $q_{e}=-e$ and mass $m_{e}$ is bound to a nucleus of effective charge $q_{N}=e$ and mass $m_{N}\gg m_{e}$. The binding interaction is modeled by a
harmonic oscillator potential, $V^{HO}=\frac{\mu}{2}(\omega_{x}^{2}x^{2}+\omega_{y}^{2}y^{2}+\omega_{z}^{2}z^{2})$, 
to which we add a term $V_{C}=C\:xyz$ to break the mirror symmetry perturbatively at first order. The coordinates $x$, $y$, $z$ are those of the relative position vector, 
$\mathbf{r}=\mathbf{r}_{N}-\mathbf{r}_{e}$, and $\mu=\frac{m_{N}m_{e}}{M}$ with $M=m_{N}+m_{e}$. The center of mass position vector is 
$\mathbf{R}=(m_{N}\mathbf{r}_{N}+m_{e}\mathbf{r}_{e})/M$. $V_{C}$ 
was firstly introduced by Condon \emph{et al}. \cite{Condon,Condon2} to explain the rotatory power of chiral compounds with a single oscillator model. 
The parameter $C$ is a  pseudoscalar which stems from the Coulomb interaction of the oscillator with its chiral environment.
We use the same model to calculate the polarizability and the rotatory power
of the molecule. Thus, we can express our result for $\mathbf{P}^{\textrm{Cas}}$ in these quantities, and
obtain a quantitative estimate based on the experimental data available for the rotatory power and refractive index of an actual compound.  
When an external uniform and constant magnetic field
$\mathbf{B}_{0}$ is applied the total Hamiltonian of the system reads, in the Coulomb gauge, $H=H_{0}+H_{EM}+W$, with
\begin{align}
H_{0}&=\sum_{i=e,N}\frac{1}{2m_{i}}[\mathbf{p}_{i}-q_{i}\mathbf{A}_{0}(\mathbf{r}_{i})]^{2}+V^{HO}+V_{C},\label{H0}\\
H_{EM}&=\sum_{\mathbf{k},\mathbf{\epsilon}}\hbar\omega_{\mathbf{k}}(a^{\dagger}_{\mathbf{k}\mathbf{\epsilon}}a_{\mathbf{k}\mathbf{\epsilon}}+\frac{1}{2})
+\frac{1}{2\mu_{0}}\int\textrm{d}^{3}r\mathbf{B}_{0}^{2},\label{HEM}\\
W&=\sum_{i=e,N}\frac{-q_{i}}{m_{i}}[\mathbf{p}_{i}-q_{i}\mathbf{A}_{0}(\mathbf{r}_{i})]\cdot\mathbf{A}(\mathbf{r}_{i})
 +\frac{q^{2}_{i}}{2m_{i}}\mathbf{A}^{2}(\mathbf{r}_{i}).\nonumber
\end{align}
In the electromagnetic (EM) vector potential we have separated the contribution of the external classical field, $\mathbf{A}_{0}(\mathbf{r}_{i})=\frac{1}{2}\mathbf{B}_{0}\wedge\mathbf{r}_{i}$,
from the one of the quantum field operator, $\mathbf{A}(\mathbf{r}_{i})$. In the sum, $a^{\dagger}_{\mathbf{k}\mathbf{\epsilon}}$ and $a_{\mathbf{k}\mathbf{\epsilon}}$
are the creation and annihilation operators of photons with momentum 
$\hbar\mathbf{k}$, frequency $\omega_{\mathbf{k}}=ck$ and polarization vector $\mathbf{\epsilon}$ respectively. 
The magnetostatic energy is a constant irrelevant to us that we will discard. 
The system obeying the above Hamiltonian posseses a conserved pseudo-momentum \cite{Kawka,Herold,Dippel}, 
\begin{equation}\label{Po}
\mathbf{K}=\mathbf{P}+\frac{e}{2}\mathbf{B}_{0}\wedge\mathbf{r}+\sum_{\mathbf{k},\mathbf{\epsilon}}\hbar\mathbf{k}(a^{\dagger}_{\mathbf{k}\mathbf{\epsilon}}a_{\mathbf{k}\mathbf{\epsilon}}+\frac{1}{2}),
\end{equation}
which satisfies $[H,\mathbf{K}]=\mathbf{0}$. Its eigenvalues are therefore good quantum numbers. $\mathbf{K}$ differs from both the conjugate 
total momentum, $\mathbf{P}=\mathbf{p}_{e}+\mathbf{p}_{N}$, 
and the kinetic momentum, 
$\mathbf{P}_{\mathrm{kin}}=M\dot{\mathbf{R}}=\mathbf{P}-\frac{e}{2}\mathbf{B}_{0}\wedge\mathbf{r}-e[\mathbf{A}(\mathbf{r}_{N})-\mathbf{A}(\mathbf{r}_{e})]$.
The terms in $\mathbf{K}$ can be also arranged as,
\begin{equation}\label{P}
\mathbf{K}=\mathbf{P}_{\textrm{kin}}+\mathbf{P}_{\textrm{Abr}}+\mathbf{P}^{\textrm{Cas}}_{\parallel}+\mathbf{P}^{\textrm{Cas}}_{\perp},
\end{equation}
where, apart from the kinetic momentum, $\mathbf{P}_{\textrm{Abr}}=e\mathbf{B}_{0}\wedge\mathbf{r}$ is the Abraham momentum and we define the Casimir momentum, 
$\mathbf{P}^{\textrm{Cas}}=\mathbf{P}^{\textrm{Cas}}_{\parallel}+\mathbf{P}^{\textrm{Cas}}_{\perp}$, as the 
momentum of the vacuum field. $\mathbf{P}^{\textrm{Cas}}$ is composed of a longitudinal Casimir momentum, 
$\mathbf{P}^{\textrm{Cas}}_{\parallel}=e[\mathbf{A}(\mathbf{r}_{N})-\mathbf{A}(\mathbf{r}_{e})]$, and a transverse Casimir momentum, 
$\mathbf{P}^{\textrm{Cas}}_{\perp}=\sum_{\mathbf{k},\mathbf{\epsilon}}\hbar\mathbf{k}(a^{\dagger}_{\mathbf{k}\mathbf{\epsilon}}a_{\mathbf{k}\mathbf{\epsilon}}+\frac{1}{2})$
\cite{Cohen}. Note that in the Coulomb gauge $\mathbf{A}$ is fully transverse.\\
\indent Under the action of a time-varying magnetic field, $\mathbf{B}_{0}(t)$, the time derivative of the expectation value of $\mathbf{K}$
in the ground state, $\langle\mathbf{K}\rangle$, vanishes,
\begin{equation}
\frac{\textrm{d}\langle\mathbf{K}\rangle}{\textrm{d}t}=i\hbar^{-1}\langle[H,\mathbf{K}]\rangle+e\frac{\partial\mathbf{B}_{0}(t)}{\partial t}\wedge\langle\mathbf{r}\rangle=\mathbf{0}.\nonumber
\end{equation}
This follows from $[H,\mathbf{K}]=\mathbf{0}$ and $\langle\mathbf{r}\rangle=\mathbf{0}$ for the chiral but unpolarized ground state \cite{Note2}. 
The latter ensures also that the variation of the Abraham momentum vanishes in the ground state. From the expectation value of Eq.(\ref{P}) this implies that,
for an arbitrary variation of the magnetic field, the variation of the kinetic momentum of the chiral oscillator is equivalent in magnitude and opposite 
in sign to the variation of the Casimir momentum of the vacuum field,
\begin{equation}\label{dPs}
\delta\langle\mathbf{P}_{\textrm{kin}}\rangle=-\delta\langle\mathbf{P}^{\textrm{Cas}}\rangle. 
\end{equation}
\indent Let us consider the molecule initially at rest in its ground state at zero magnetic field. During the switching of the magnetic field
Eq.(\ref{dPs}) implies that $\langle\mathbf{P}_{\textrm{kin}}\rangle=-\langle\mathbf{P}^{\textrm{Cas}}\rangle$ at any time. This relation makes 
the magnetochiral Casimir momentum an observable quantity. In particular this equality 
holds well after the switching has ended and the magnetic field achieves a stationary value $\mathbf{B}_{0}$. In the following we 
evaluate the Casimir momentum in the asymptotically stationary situation in which the molecule is in its ground state at constant kinetic momentum and 
constant magnetic field $\mathbf{B}_{0}$.\\
\indent In the first stage of the calculation we neglect the vacuum field. The conserved pseudo-momentum of the resultant Hamiltonian, $H_{0}$, is 
$\mathbf{K}_{0}=\mathbf{P}_{\textrm{kin}}+e\mathbf{B}_{0}\wedge\mathbf{r}$, with $[H_{0},\mathbf{K}_{0}]=\mathbf{0}$ and continuous eigenvalues  
 $\mathbf{Q}$. The unitary operator 
U$=\exp{[i(\mathbf{Q}-\frac{e}{2}\mathbf{B}_{0}\wedge\mathbf{r})\cdot\mathbf{R}/\hbar]}$ maps the Hamiltonian $H_{0}$ into
$\tilde{H}_{0}=$U$^{\dagger}H_{0}$U, which conveniently separates internal and external motion \cite{Herold,Dippel},
\begin{equation}
\tilde{H}_{0}=\frac{1}{2M}\mathbf{Q}^{2}+\frac{1}{2\mu}\mathbf{p}^{2}+V^{HO}+V_{C}+V_{Z}+\mathcal{O}(\textrm{Q}\textrm{B}_{0},\textrm{B}_{0}^{2}).\label{effective}
\end{equation}
In this equation $\mathbf{p}$ is the conjugate momentum of $\mathbf{r}$, 
$\mathbf{p}=\mu(\mathbf{p}_{N}/m_{N}-\mathbf{p}_{e}/m_{e})$, and $V_{Z}=\frac{e}{2\mu^{*}}(\mathbf{r}\wedge\mathbf{p})\cdot\mathbf{B}_{0}$ is the Zeeman potential with $\mu^{*}=\frac{m_{N}m_{e}}{m_{N}-m_{e}}$. 
 
The ground state of the Hamiltonian $\tilde{H}_{0}$ is, up to order $C$B$_{0}$ in stationary perturbation theory, 
%\begin{eqnarray}
%|\tilde{\Omega}_{0}\rangle&=&|0\rangle-\mathcal{C}|111\rangle
%-i\mathcal{B}_{0}^{x}\eta^{zy}|011\rangle\nonumber\\
%&+&i\mathcal{B}_{0}^{x}\mathcal{C}\eta^{zy}(|100\rangle+2|122\rangle)\nonumber\\
%&-&\sqrt{2}i\mathcal{B}_{0}^{x}\mathcal{C}\bigl(\frac{2\omega_{y}-\omega_{x}\eta^{zy}}{\omega_{x}+2\omega_{y}}|120\rangle
%-\frac{2\omega_{z}+\omega_{x}\eta^{zy}}{\omega_{x}+2\omega_{z}}|102\rangle\bigr)\nonumber\\
%&+&\textrm{ cyclic permutations}.\nonumber
%\end{eqnarray}
\begin{eqnarray}
|\tilde{\Omega}_{0}\rangle&=&|0\rangle-\mathcal{C}|111\rangle
-i\mathcal{B}_{0}^{z}\eta^{yx}|110\rangle\nonumber\\
&+&i\mathcal{B}_{0}^{z}\mathcal{C}\eta^{yx}\left(|001\rangle+2|221\rangle\right)\nonumber\\
&-&\sqrt{2}i\mathcal{B}_{0}^{z}\mathcal{C}\left(\frac{2\omega_{x}-\omega_{z}\eta^{yx}}{\omega_{z}+2\omega_{x}}|201\rangle
-\frac{2\omega_{y}+\omega_{z}\eta^{yx}}{\omega_{z}+2\omega_{y}}|021\rangle\right)\nonumber\\
&+&\sum\textrm{ cyclic permutations}.\label{vacuum}
\end{eqnarray}
Correspondingly, the ground state of $H_{0}$ is $|\Omega_{0}\rangle=$U$|\tilde{\Omega}_{0}\rangle$, with a pseudo-momentum $\mathbf{Q}_{0}$ to be fixed. 
The fact that $\langle\Omega_{0}|\mathbf{r}|\Omega_{0}\rangle=\mathbf{0}$ implies that $\mathbf{Q}_{0}$ is the kinetic momentum of the oscillator, 
i.e., no other contribution exists to the pseudomomentum $\langle\mathbf{K}_{0}\rangle$ of the bare molecule.
In the above equation the states $|n_{x} n_{y} n_{z}\rangle$ refer to the eigenstates of the harmonic oscillator Hamiltonian. The dimensionless parameters are, $\mathcal{B}_{0}^{i}=\frac{eB_{0}^{i}}{4\mu^{*}\sqrt{\omega_{j}\omega_{k}}}$ with 
$i\neq j\neq k$ and $i\neq k$, $\mathcal{C}=\frac{C\hbar^{1/2}}{(2\mu)^{3/2}(\omega_{x}+\omega_{y}+\omega_{z})(\omega_{x}\omega_{y}\omega_{z})^{1/2}}$, 
$\eta^{ij}=\frac{\omega_{i}-\omega_{j}}{\omega_{i}+\omega_{j}}$. Here, the indices $i,j,k$ take on the three spatial directions, $x,y,z$. 
The $\eta$ factors are assumed to be small quantities which quantify the anisotropy of the oscillator. 
They \emph{all} have to be nonzero for the optical activity of the molecule to survive rotational averaging. In the following, all our calculations restrict to  
the lowest order in $\mathcal{C}$, $\mathcal{B}_{0}^{i}$ and $\eta^{ij}$.\\
\indent Next, we couple the oscillator to the vacuum field and evaluate the Casimir momentum, $\langle\mathbf{P}^{\textrm{Cas}}\rangle=\langle\Omega|\mathbf{P}^{\textrm{Cas}}|\Omega\rangle$.
Here, $|\Omega\rangle$ is the ground state of the oscillator when coupled to the vacuum field. It will be computed applying up to second order perturbation 
theory to $|\Omega_{0}\rangle$ with the interaction potential $W$. Our calculations show that, for nonrelativistic velocities, terms which depend
on the kinetic momentum, $\mathbf{Q}_{0}$, are negligible in the computation of $\langle\mathbf{P}^{\textrm{Cas}}\rangle$ except for divergent mass renormalization
terms that can be made disappear \cite{SM,Kawka}.
Hereafter we will take $\mathbf{Q}_{0}=\mathbf{0}$ for simplicity without loss of generality. The transverse momentum, 
$\langle\mathbf{P}_{\perp}^{\textrm{Cas}}\rangle=\sum_{\mathbf{k},\mathbf{\epsilon}}\hbar\mathbf{k}\langle\Omega|a^{\dagger}_{\mathbf{k}\mathbf{\epsilon}}a_{\mathbf{k}\mathbf{\epsilon}}|\Omega\rangle$,
is found to be finite but small. The longitudinal momentum, 
$\langle\mathbf{P}_{\parallel}^{\textrm{Cas}}\rangle=e\langle\Omega|\mathbf{A}(\mathbf{r}_{N})-\mathbf{A}(\mathbf{r}_{e})|\Omega\rangle$, gives the dominant contribution 
that we will evaluate in detail first.
%\begin{figure}[h]
%\includegraphics[height=5.2cm,width=5.9cm,clip]{Feynman.eps}
%\caption{Feynman diagrams of the process contributing to $\langle\mathbf{P}^{\textrm{Cas}}_{\parallel}\rangle$ in Eq.(\ref{long}).
%$(a)$ Dominant processes. $(b)$ Subdominant processes.}
%\end{figure}\label{Feynman}

At $\mathcal{O}(C$B$_{0})$ and lowest order in the coupling constant $e$, it suffices to 
compute $|\Omega\rangle$ applying first order perturbation theory to $|\Omega_{0}\rangle$. 
We use the U-transformed states and the U-transformed potential, with U$=\exp{[-i\frac{e}{2\hbar}(\mathbf{B}_{0}\wedge\mathbf{r})\cdot\mathbf{R}]}$,
\begin{eqnarray}\label{tW}
\tilde{W}&=&-\frac{e}{m_{N}}\left(\mathbf{p}+\frac{m_{N}}{M} \mathbf{P}-\frac{e}{2}\mathbf{B}_{0}\wedge\mathbf{r}\right)\cdot
\mathbf{A}(\mathbf{R}+\frac{m_{e}}{M}\mathbf{r})\nonumber\\
&-&\frac{e}{m_{e}}\left(\mathbf{p}-\frac{m_{e}}{M} \mathbf{P}+\frac{e}{2}\mathbf{B}_{0}\wedge\mathbf{r}\right)\cdot
\mathbf{A}(\mathbf{R}-\frac{m_{N}}{M}\mathbf{r})\nonumber\\
&+&\frac{e^{2}}{2m_{N}}\textrm{A}^{2}(\mathbf{R}+\frac{m_{e}}{M}\mathbf{r})
+\frac{e^{2}}{2m_{e}}\textrm{A}^{2}(\mathbf{R}-\frac{m_{N}}{M}\mathbf{r}),
\end{eqnarray}
to arrive at,
\begin{equation}
\langle\mathbf{P}^{\textrm{Cas}}_{\parallel}\rangle=\sum_{\mathbf{Q},I,\gamma_{\mathbf{k}\mathbf{\epsilon}}}
\frac{\langle\tilde{\Omega}_{0}|e\Delta\mathbf{A}|\mathbf{Q},I,\gamma\rangle
\langle\mathbf{Q},I,\gamma|\tilde{W}|\tilde{\Omega}_{0}\rangle}{E_{0}-E_{Q,I,k}}+c.c.,\label{pert1} 
\end{equation}
where $\Delta\mathbf{A}=\mathbf{A}(\mathbf{r}_{N})-\mathbf{A}(\mathbf{r}_{e})$, $E_{0}=\hbar(\omega_{x}+\omega_{y}+\omega_{z})/2$ and $E_{Q,I,k}$ is 
the energy of intermediate states, $|\mathbf{Q},I,\gamma\rangle=|\mathbf{Q},I\rangle\otimes|\gamma_{\mathbf{k}\mathbf{\epsilon}}\rangle$. 
The atomic states $|\mathbf{Q},I\rangle$ are eigenstates of $\tilde{H}_{0}$ and may have a priori any pseudo-momentum $\mathbf{Q}$. The EM states, 
$|\gamma_{\mathbf{k}\mathbf{\epsilon}}\rangle$, are 1-photon states with momentum $\hbar\mathbf{k}$ and polarization vector $\mathbf{\epsilon}$.
Zeros in the denominator are avoided in the summation. It turns out that only the terms proportional to $\mathbf{p}\cdot\mathbf{A}$ 
in $\tilde{W}$ yield a nonvanishing contribution to $\langle\mathbf{P}^{\textrm{Cas}}_{\parallel}\rangle$. Writing the EM quantum field in Eq.(\ref{pert1}) as 
usual \cite{Loudon}, 
\begin{equation}
\mathbf{A}(\mathbf{r})=\sum_{\mathbf{k},\mathbf{\epsilon}}\sqrt{\frac{\hbar}{2ck\mathcal{V}\epsilon_{0}}}
[\mathbf{\epsilon}a_{\mathbf{k}}e^{i\mathbf{k}\cdot\mathbf{r}}+\mathbf{\epsilon}^{*}a^{\dagger}_{\mathbf{k}}e^{-i\mathbf{k}\cdot\mathbf{r}}],
\end{equation}
with $\mathcal{V}$ a generic volume, $c$ the speed of light and $\epsilon_{0}$ the vacuum permittivity, passing the sums over $\mathbf{Q}$ and $\mathbf{k}$ to continuum integrals and by summing over polarization states we arrive at,
\begin{eqnarray}\label{long}
\langle\mathbf{P}^{\textrm{Cas}}_{\parallel}\rangle&=&\frac{\hbar e^{2}}{2c\epsilon_{0}}\int\frac{\textrm{d}^{3}k}{(2\pi)^{3}k}\langle\tilde{\Omega}_{0}|
(e^{i\frac{m_{e}}{M}\mathbf{k}\cdot\mathbf{r}}-e^{-i\frac{m_{N}}{M}\mathbf{k}\cdot\mathbf{r}})\nonumber\\
&\times&\frac{(\mathbb{I}-\frac{\mathbf{k}\otimes\mathbf{k}}{k^{2}})\cdot}{\hbar^{2}k^{2}/2M+\hbar ck-E_{0}+\tilde{H}_{0}}\nonumber\\
&\times&\left(\frac{e^{-i\frac{m_{e}}{M}\mathbf{k}\cdot\mathbf{r}}}{m_{N}}+\frac{e^{i\frac{m_{N}}{M}\mathbf{k}\cdot\mathbf{r}}}{m_{e}}\right)\mathbf{p}|\tilde{\Omega}_{0}\rangle+c.c.,
\end{eqnarray}
where $\mathbb{I}$ is the unit dyadic and the kinetic energy $\hbar^{2}k^{2}/2M$ in the denominator is a consequence of the momentum recoil of photon to atom, 
$\mathbf{Q}=-\hbar\mathbf{k}$. When the oscillatory exponential operators on the left of the 
Hamiltonian operator are  moved to the right, two types of term appear. To the first type belong those terms in which the products of the oscillatory exponentials cancel. 
They correspond to processes in which a photon is created and annihilated at the same particle (i.e., electron or nucleus). The 
second class contains terms in which the product of the exponentials combine to $e^{\pm i\mathbf{k}\cdot\mathbf{r}}$. 
They correspond to processes in which a photon is created and annihilated 
at different particles. The first category dominates, while the second category is a factor  
$\sqrt{E_{0}/\mu c^{2}}$ smaller and will be neglected.

The dominant terms are,
\begin{eqnarray}
\langle\mathbf{P}^{\textrm{Cas}}_{\parallel}\rangle&=&\Re\Bigl\{\Bigl[\frac{\hbar e^{2}}{3\pi^{2}c\epsilon_{0}m_{e}}\int_{0}^{\infty}\langle \tilde{\Omega}_{0}|
\frac{k\textrm{d}k}{\frac{\hbar^{2}k^{2}}{2m_e}+\hbar ck-E_{0}+H^{HO}}\nonumber\\
&\times&(V_{C}+V_{Z})\frac{1}{\frac{\hbar^{2}k^{2}}{2m_e}+\hbar ck-E_{0}+H^{HO}}\mathbf{p}|\tilde{\Omega}_{0}\rangle
\nonumber\\
&-&\frac{\hbar e^{2}}{3\pi^{2}c\epsilon_{0}m_{e}}\int_{0}^{\infty}\langle 0|
\frac{k\textrm{d}k}{\frac{\hbar^{2}k^{2}}{2m_e}+\hbar ck-E_{0}+H^{HO}}\nonumber\\
&\times&(V_{C}+V_{Z})\frac{1}{\frac{\hbar^{2}k^{2}}{2m_e}+\hbar ck-E_{0}+H^{HO}}\nonumber\\
&\times&(V_{Z}+V_{C})\frac{1}{\frac{\hbar^{2}k^{2}}{2m_e}+\hbar ck-E_{0}+H^{HO}}\mathbf{p}|0\rangle\Bigr]\nonumber\\
&-&[m_{e}\leftrightarrow m_{N}]\Bigr\},\label{Ppar3}
\end{eqnarray}
where $-[m_{e}\leftrightarrow m_{N}]$ means that the same expression within the square brackets must be evaluated exchanging 
$m_{e}$ with $m_{N}$ and subtracted. We write $H^{HO}=$p$^{2}/2\mu+V^{HO}$. 
Note that, since both the chiral interaction $V_{C}$ and the Zeeman effect $V_{Z}$ featuring in $\tilde{H}_{0}$ are treated as perturbations to 
the harmonic oscillator, the denominator in Eq.(\ref{long}) is expanded up to order $V_{C}V_{Z}$ in Eq.(\ref{Ppar3}). 
Hence, only terms at $\mathcal{O}(C$B$_{0})$ must be retained. 
When moving the exponential operators from the left to the right of 
the Hamiltonian operator in Eq.(\ref{long}), respecting the canonical commutation relations the momentum $\mathbf{p}$ in $\tilde{H}_{0}$ gets shifted,
$\mathbf{p}\rightarrow\mathbf{p}\pm \frac{m_{N,e}}{M}\hbar\mathbf{k}$. As a result, the kinetic energy in the denominators of Eq.(\ref{Ppar3}) 
becomes $\hbar^{2}k^{2}/2m_{e,N}$ and an additional term $\pm \hbar\mathbf{k}\cdot\mathbf{p}/m_{e,N}$ shows up there. The latter is a Doppler shift term which 
has been neglected in Eq.(\ref{Ppar3}). 

No divergent terms appear in Eq.(\ref{Ppar3}) for $\langle\tilde{\Omega}_{0}|\mathbf{p}|\tilde{\Omega}_{0}\rangle=\mathbf{0}$ \cite{SM}, and all the integrals 
are convergent at $\mathcal{O}(C$B$_{0})$. When integrating over $k$ up to $\infty$ the leading contribution is logarithmic. 
The end result can be expressed as,
%\begin{eqnarray}\label{Pkinparx}
%\langle \mathbf{P}^{\textrm{Cas}}_{\parallel}\rangle&=&\frac{Ae^{3}\ln{(m_{e}/m_{N})}}{48\pi^{2}c\epsilon_{0}\mu\mu^{*}
%(\omega_{x}+\omega_{y}+\omega_{z})}\nonumber\\
%&\times&\bigl(\frac{\eta^{zy}\textrm{B}_{0}^{x}}{\omega_{y}\omega_{z}}\hat{x}+\frac{\eta^{xz}\textrm{B}_{0}^{y}}{\omega_{x}\omega_{z}}\hat{y}
%+\frac{\eta^{yx}\textrm{B}_{0}^{z}}{\omega_{x}\omega_{y}}\hat{z}\bigr).
%\end{eqnarray}
\begin{equation}\label{Pkinparx}
\langle \mathbf{P}^{\textrm{Cas}}_{\parallel}\rangle=\frac{Ce^{3}\ln{(m_{N}/m_{e})}}{96\pi^{2}c\epsilon_{0}\mu\mu^{*}
(\omega_{x}+\omega_{y}+\omega_{z})}\sum_{i,j,k}\varepsilon_{ijk}\frac{\textrm{B}_{0}^{i}\eta^{kj}}{\omega_{k}\omega_{j}}\hat{i},
\end{equation}
where $\varepsilon_{ijk}$ is the three-dimensional Levi-Civita tensor and $\hat{i}$ is a unitary vector along the $i$ axis. 
Note that $\eta^{kj}= -\eta^{jk}$. We thus need full anisotropy of the oscillator quantified by $\eta^{jk}$ to make the Casimir momentum nonzero.
The momentum is still not averaged over the random orientation of the chiral object. To do so we must project $\langle \mathbf{P}^{\textrm{Cas}}\rangle$ 
onto the magnetic field vector and extract the trace,
\begin{equation}\label{PkinparAverage}
\langle\mathbf{P}^{\textrm{Cas}}_{\parallel}\rangle_{rot}=\frac{Ce^{3}\ln{(m_{e}/m_{N})}\mathbf{B}_{0}}{144\pi^{2}c\epsilon_{0}\mu\mu^{*}\omega_{x}\omega_{y}\omega_{z}}
\eta^{zy}\eta^{xz}\eta^{yx},
\end{equation}
where the subscript $rot$ denotes rotational average. The logarithmic dependence on Eqs.(\ref{Pkinparx},\ref{PkinparAverage}) may be disputed since
it results from wave numbers larger than $m_{N,e}c/\hbar$. It is possible that relativistic corrections are needed, which lie outside 
the scope of the present Letter. 

The transverse Casimir momentum, $\langle\mathbf{P}^{\textrm{Cas}}_{\perp}\rangle$, corresponds to the expectation value of the operators in the sum over modes of Eq.(\ref{P}). Its evaluation involves a second order 
pertubation calculation. %whose details will be published somewhere else \cite{Prep}. 
The end result is approximately,
\begin{align}
\langle\mathbf{P}^{\textrm{Cas}}_{\perp}\rangle_{rot}&=
\frac{-1.06Ce^{3}\mathbf{B}_{0}}{144\pi^{2}c\epsilon_{0}m_{e}^{2}\omega_{x}\omega_{y}\omega_{z}}
\eta^{zy}\eta^{xz}\eta^{yx}\label{Pperptot}\\
&\simeq\langle\mathbf{P}^{\textrm{Cas}}_{\parallel}\rangle_{rot}/\ln{(m_{N}/m_{e})}.\nonumber
\end{align} 
For instance, if we take for $m_{N}$ the mass of a carbon atom, $\langle\mathbf{P}^{\textrm{Cas}}_{\perp}\rangle_{rot}$ is an order of magnitude less than  
$\langle\mathbf{P}^{\textrm{Cas}}_{\parallel}\rangle_{rot}$.

Eqs.(\ref{PkinparAverage}) and (\ref{Pperptot}) give simple formulas for $\langle\mathbf{P}^{\textrm{Cas}}_{\parallel}\rangle_{rot}$ and $\langle\mathbf{P}^{\textrm{Cas}}_{\perp}\rangle_{rot}$
in terms of the chiral parameter $C$, the magnetic field and the natural frequencies of the oscillator. We will now express them 
in terms of the static optical rotatory power and electric polarizability of the molecule, which can be easily calculated for this simple model and which are both directly observable.
To this end we derive an equation for the electric dipole moment induced in the state $|\Omega_{0}\rangle$ of the molecule by a probe EM field of frequency $\omega$
whose electric field is $\mathbf{E}_{\omega}$. For this we apply first order perturbation theory to $|\tilde{\Omega}_{0}\rangle$ 
using the potential $\tilde{W}$ in which the 
quantum field must be substituted by a classical EM vector potential from which $\mathbf{E}_{\omega}$ derives. Expanding $\tilde{W}$ up to electric quadrupole terms
and performing the rotational averages we can write the molecule's electric dipole moment in SI units as \cite{Prep},
\begin{equation}
\langle\mathbf{d}\rangle_{rot}=\alpha_{E}\mathbf{E}_{\omega}+\chi\:\mathbf{B}_{0}\wedge\dot{\mathbf{E}}_{\omega}+i\beta\:\mathbf{k}\wedge\mathbf{E}_{\omega}
+\gamma\:(\mathbf{k}\cdot\mathbf{B}_{0})\mathbf{E}_{\omega}.\label{dm}\\
\end{equation}
In this equation $\alpha_{E}$ is the electric polarizability, $\chi$ describes the Faraday effect, $\beta$ is the molecular rotatory factor responsible for 
the natural optical activity and $\gamma$ gives rise to the magneto-chiral 
anisotropy. All polarizabilities tend to constant values in the static limit, $\omega\rightarrow0$, where we find \cite{Prep},
\begin{equation}
 \beta(0)/ \alpha_{E}(0)=\frac{\hbar C\mathcal{D}}{8\mu\mu^{*}\omega_{x}\omega_{y}\omega_{z}}\eta^{zy}\eta^{xz}\eta^{yx}.\label{2em}
\end{equation}
In this formula $\mathcal{D}$ is a rational polynomial of the frequencies of the oscillator which approximates $\mathcal{D}\approx1$ for not too large 
anisotropy. By comparing the above expression with Eqs.(\ref{PkinparAverage},\ref{Pperptot}) we conclude that 
the total Casimir momentum, $\langle\mathbf{P}^{\textrm{Cas}}\rangle_{rot}=\langle\mathbf{P}^{\textrm{Cas}}_{\parallel}\rangle_{rot}+\langle\mathbf{P}^{\textrm{Cas}}_{\perp}\rangle_{rot}$,
can be written approximately as,
\begin{equation}\label{main1}
\langle\mathbf{P}^{\textrm{Cas}}\rangle_{rot}\simeq\frac{-2\alpha}{9\pi}\frac{\beta(0)}{\alpha_{E}(0)}[\ln{(m_{N}/m_{e})}+1]e\mathbf{B}_{0},
\end{equation}
where $\alpha$ is the fine structure constant. Thus, we confirm our conjecture, $\mathbf{P}^{\textrm{Cas}}= g e\mathbf{B}_0$, with
$g=\frac{-2\alpha}{9\pi}\frac{\beta(0)}{\alpha_{E}(0)}[\ln{(m_{N}/m_{e})}+1]$.
We speculate that, apart from constants of order unity, this expression is model-independent. %Note that the Casimir momentum computed here corresponds to the kinetic 
%momentum --with opposite sign- transferred to a chiral molecule initially at rest, during the adiabatic switching of the external magnetic field from zero up to its 
%final value, $\mathbf{B}_{0}$.

The presence of $\hbar$ in Eq.(\ref{2em}) as well as its absence in Eqs.(\ref{Pkinparx},\ref{Pperptot}) deserve a comment since the polarizabilities
$\alpha_{E}$ and $\beta$ are classical observables. The fact that $\beta$ is proportional to $\hbar$ 
is a consequence of quantum mechanics. For a given set of natural frequencies, all the atomic lengths in the problem are determined by quantum mechanics. In 
particular, $\beta(0)/ \alpha_{E}(0)$ is a length, a fraction of the electronic Compton wavelength, that we identify with a chiral length, $l_{ch}$ 
characteristic of the non-locality of the optical response. Therefore, 
we can write $\textrm{P}^{\textrm{Cas}}\sim\alpha\:e$B$_{0}l_{ch}$ and thus interpret it as the leading QED correction to the classical Abraham momentum.

%Next, we compare our result with earlier work on Casimir momentum wich is based on the application of the fluctuation-dissipation theorem
%and worked out in the dipole approximation \cite{Bart2008,Babington}. For our model that approach yields,
%\begin{equation}\label{laPC2}
%\mathbf{P}^{\textrm{Cas}}=\frac{-\hbar\mathbf{B}_{0}}{3\pi^{2}\epsilon_{0}}\Re\int_{0}^{k_{max}}\textrm{d}k\:k^{4}\gamma(k).
%\end{equation}
%This expression is analogous to the formula obtained in Refs.~\cite{Feigel,Croze} for the transverse Casimir momentum transferred to a homogeneous magnetoelectric medium.
%In contrast to the quantum calculation, the integral of Eq.(\ref{laPC2}) diverges quadratically for $k_{max}\rightarrow\infty$ and needs a UV cut-off to yield a finite result. 
%Upon comparing the result of this integral with the quantum result of Eq.(\ref{Pperptot}) we have identified this cut-off with the inverse Compton wavelength,
%$k_{max}\sim\frac{\mu c}{\hbar}$. The details of the calculation will be published elsewhere \cite{Prep}. 

Finally, we estimate the orders of magnitude of the Casimir momentum transferred to a single molecule of an actual chiral compound. 
To this aim we derive values for $\beta(0)$ and $\alpha_{E}(0)$ from experimental parameters. 
Let us consider a molecular medium with a number density $\rho$. 
At leading order in $\rho$ and far from resonances the rotatory power of the medium, $\varphi$, relates to the molecular rotatory factor $\beta$ as, 
$\varphi=4\pi^{2}\rho\beta(0)/\lambda^{2}\epsilon_{0}$, and the refractive index is $n= 1+\rho\alpha_{E}(0)/2\epsilon_{0}$. 
As an example we take the chiral compound 2-octanol, C$_{8}$H$_{18}$O. According to Table I
of Ref.~\cite{Condon},  at $\lambda=4800$\AA{}, $n=1.43$ and its specific rotatory power is [$\varphi$]$^{4800}$=15.46 deg/dm/(g/cm$^{3}$). Its mass density is 
$0.82$ g/cm$^{3}$ and its molecular mass is 130.2 u. We find, $\beta(0)=3\times10^{-53}$radC$^{2}$m$^{3}/$J, 
$\alpha_{E}(0)=2\times10^{-39}$m$^{2}$C$^{2}/$J. %$\rho\approx3.8\times10^{27}$m$^{-3}$, \varphi^{4800}\simeq2.2rad/m.
 Considering that the chiral center is one of the carbon atoms and using B$_{0}=10$T, we find 
$\langle \textrm{P}^{\textrm{Cas}}\rangle_{rot}= 1.4\times10^{-34}$\:kg$\:$m/s, which lies within the scope of current experimental
observations \cite{Rikken1,Rikken2}. For the molecule considered, this momentum corresponds to a velocity of $0.6$ nm/s.

In this Letter we have carried out a quantum computation of the kinetic Casimir momentum transferred from the quantum vacuum to a chiral oscillator subject to an external magnetic 
field. The result is free of divergences, linear in $\mathbf{B}_{0}$ and proportional to $\alpha$, which proves the genuine QED nature of the effect. 
Our calculation is based on a simplified molecular model which allows us to express the final result in 
terms of observable parameters, i.e., the molecular rotatory factor and the electric polarizability. Eq.(\ref{main1}) is the main result of this Letter. 
A quantitative prediction has been made whose verification is accessible to current experiments.

This work was supported by the ANR contract PHOTONIMPULS
ANR-09-BLAN-0088-01.

\end{document}